\documentclass[preprint2]{aastex631}

\usepackage{graphicx,bm,amsmath}
\usepackage{braket}
\usepackage{multirow}

\newcommand{\rmd}{\mathrm{d}}
\newcommand{\rme}{\mathrm{e}}
\newcommand{\rmi}{\mathrm{i}}
\renewcommand{\vec}[1]{\boldsymbol #1}
\newcommand{\abs}[1]{\left|#1\right|}
\newcommand{\sumint}%
{\mathop{\hbox{$\displaystyle\sum\kern-13.2pt\int\kern1.5pt$}}}

\shorttitle{Vibrationally resolved photoionization of H$_{2}^{+}$}
\shortauthors{Singor et al.}
\graphicspath{{./}{figures/}}

\begin{document}

\title{Photoionization from the ground and excited vibrational states of H$_{2}^{+}$ and its deuterated isotopologues}

\author[0000-0002-9098-9136]{Adam Singor}
\affiliation{Department of Physics and Astronomy, Curtin University, Perth, Western Australia 6102, Australia}

\author[0000-0002-9900-9712]{Liam H. Scarlett}
\affiliation{Department of Physics and Astronomy, Curtin University, Perth, Western Australia 6102, Australia}

\author[0000-0003-0473-379X]{Mark C. Zammit}
\affiliation{Theoretical Division, Los Alamos National Laboratory, Los Alamos, NM 87545, United States of America}

\author[0000-0001-7554-8044]{Igor Bray}
\affiliation{Department of Physics and Astronomy, Curtin University, Perth, Western Australia 6102, Australia}

\author[0000-0002-3951-9016]{Dmitry V. Fursa}
\affiliation{Department of Physics and Astronomy, Curtin University, Perth, Western Australia 6102, Australia}



\begin{abstract}

Photoionization cross sections and rate coefficients have been calculated for all bound vibrational levels of the 1s$\sigma_{\mathrm{g}}$ state of H$_{2}^{+}$, HD$^{+}$, and D$_{2}^{+}$.
The Born--Oppenheimer approximation is employed in our calculation of vibrationally--resolved photoionization cross sections.
Vibrationally--resolved and local thermal equilibrium photoionization rate coefficients are presented for photon temperatures less than $50\,000$ K and are found to be several orders of magnitude larger than previous results in the literature.
Analytic fits for the vibrationally--resolved and local thermal equilibrium photoionization rate coefficients are provided.
Near threshold oscillations in the vibrationally--resolved photoionization are observed.
A benchmark set of photoionization cross sections are presented.
Fixed--nuclei photoionization cross sections are calculated using two--center true continuum wave functions and are verified by comparison with previous calculations and are found to be in excellent agreement in all cases.
Data files for our set of benchmark cross sections, rate coefficients, and fitting parameters for H$_{2}^{+}$, HD$^{+}$, and D$_{2}^{+}$ are available on Zenodo under an open-source 
Creative Commons Attribution license: 
\dataset[doi:10.5281/zenodo.8304060]{https://doi.org/10.5281/zenodo.8304060}.

\end{abstract}



\section{Introduction}
\label{sec:intro}
Atomic and molecular hydrogen are abundant in the interstellar medium with H$_{2}^{+}$ being formed by radiative association of protons and atomic hydrogen, and ionization of H$_{2}$~\citep{Dalgarno1985,Mukherjee2006}.
Photoionization of H$_{2}^{+}$:
\begin{equation}
\gamma(E_{\gamma}) + \mathrm{H}_{2}^{+}(1s\sigma_{\mathrm{g}},v_{i}) \longrightarrow 2\mathrm{H}^{+} + e^{-}(E_{k}),
\end{equation}
is of interest in astrophysics and plasma physics~\citep{Langhoff1985,Galli1998,Heays2017}, particularly for the cooling rates of astrophysical plasma~\citep{Wiersma2009}, the composition of stellar and planetary atmospheres~\citep{Gruntman1996,Tseng2011}, the production of protons in interstellar clouds and planetary atmospheres~\citep{Ford1975}, and the ionized regions of planetary nebulae~\citep{Aleman2004}.
Photoionization rate coefficients for H$_{2}^{+}$ have been determined by~\citet{Galli1998} using the fixed--nuclei cross sections of~\citet{Bates1968}, which do not account for vibrational motion.
In plasmas, H$_{2}^{+}$ and its isotopologues exist in a distribution of vibrational levels, and hence photoionization cross sections resolved in the initial vibrational level are required.
However, such data has not been calculated previously.

Photoionization has been well--studied within the fixed--nuclei approximation \citep{Bates1953,Cohen1966,Bates1968,Brosolo1992,Brosolo1994,Colgan2003,Fojon2006,Fernandez2007,Fernandez2009A,DellaPicca2008,DellaPicca2009,Fernandez2009B,DellaPicca2011,Haxton2013,Arkhipov2018}. 
A simple approximation proposed by \citet{Cohen1966} suggested that undulations that had been previously observed in photoionization cross sections of O$_{2}$ and N$_{2}$ could be explained by interference resulting from considering the two atoms as independent sources of photoelectrons. 
They also suggested that such undulation would be present in photoionization cross sections for H$_{2}^{+}$, and also proposed that the integrated photoionization cross section for a diatomic molecule could be approximated by a corresponding hydrogen--like atomic cross section multiplied by a modulation factor,
\begin{equation}
\sigma^{(\mathrm{PI})} = \frac{\sigma_{\mathrm{H}}^{(\mathrm{PI})}(Z_{\mathrm{eff}})}{1+S}\left(1+\frac{\sin(k_{e}R)}{k_{e}R} \right), \label{eq:CF_model}
\end{equation}
where $k_{e}$ is the photoelectron momentum in atomic units, $R$ is the internuclear separation in atomic units, $S$ is the overlap integral between the 1s hydrogenic orbitals on each nuclei, and $\sigma_{\mathrm{H}}^{(\mathrm{PI})}(Z_{\mathrm{eff}})$ is the photoionization cross section for a hydrogen--like atom with an effective charge $Z_{\mathrm{eff}}$.

We present a comprehensive set of accurate benchmark fixed--nuclei photoionization cross sections and calculate photoionization cross sections for the ground and all excited vibrational levels of the 1s$\sigma_{\mathrm{g}}$ state of H$_{2}^{+}$ and its deuterated isotopologues. 
Vibrationally--resolved and local thermal equilibrium (LTE) photoionization rate coefficients for H$_{2}^{+}$, HD$^{+}$, and D$_{2}^{+}$ have been produced.
In Section~\ref{sec:theory} we provide an overview of the H$_{2}^{+}$ isotopologue target structure and the calculation of cross sections and rate coefficients for photoionization.
A comprehensive set of benchmark fixed--nuclei photoionization cross sections are presented and compared with previous results in Section~\ref{sec:FN-ion}.
In Section~\ref{sec:vib_ion}, photoionization cross sections for all bound vibrational levels of H$_{2}^{+}$, HD$^{+}$, and D$_{2}^{+}$ are presented.
Photoionization rate coefficients are presented in Section~\ref{sec:rates}.
Conclusions and future directions are formulated in Section~\ref{sec:conclusion}.
Atomic units are used unless stated otherwise.

\section{Theory}
\label{sec:theory}
In this section, we provide a brief discussion of the target structure of the H$_{2}^{+}$ molecule and its isotopologues, and the formalism for calculating the photoionization cross sections and rate coefficients. Further details are provided in Appendix~\ref{App:structure}.

The molecular target states within the Born--Oppenheimer approximation can be written in prolate spheroidal coordinates as
\begin{equation}
\Psi_{nv}(\bm{\rho},R) = \psi_{n}(\bm{\rho};R) \nu_{nv}(R) \label{eq:BO}
\end{equation}
where $n$ specifies the electronic state, $v$ is the vibrational quantum number, $\bm{\rho}$ are the electronic coordinates, $R$ is the internuclear separation, $\psi_{n}$ is the electronic wave function, and $\nu_{nv}$ is the vibrational wave function.
Rotational motion is neglected in the current calculations.

Bound electronic states, Equation~(\ref{eq:bound_st_expansion}), are obtained by diagonalizing the unseparated electronic Hamiltonian, Equation~(\ref{eq:unsep_hamiltonian}), in a Sturmian basis for each orbital angular momentum projection $m$~\citep{Zammit2017,Scarlett2021a}.
True continuum target states are calculated using the approach given by \citet{Singor2022}.
The quasi--angular wave functions are obtained by expanding the spheroidal harmonics in Equation~(\ref{eq:angular}) as a series of spherical harmonics of the same $m$. The solution to the quasi--radial equation, Equation~(\ref{eq:radial}), is started using a power series expansion and then propagated using an Adams--Moulton predictor--corrector algorithm. An asymptotic series is then used to normalize the wave function.
Bound vibrational wave functions, Equation~(\ref{eq:vib_wf}), are generated by diagonalizing the vibrational Hamiltonian in a Sturmian basis~\citep{Scarlett2021a}.

The cross section for photoionization of an H$_{2}^{+}$ isotopologue in the initial state $\ket{\Psi_{n_{i}v_{i}}}=\ket{\psi_{n_{i}}}\!\ket{\nu_{n_{i}v_{i}}}$ by an unpolarized photon with energy $E_{\gamma}$ is, (the integral over $v_{f}$ denotes integration over the dissociative continuum for a given continuum electronic state)
\begin{align}
\sigma_{n_{i}v_{i}}^{(\mathrm{PI})}(E_{\gamma}) &= \sigma_{\mathrm{T}}\frac{\pi c^{3}E_{\gamma}}{2} \sum_{\lambda_{f}} \int\rmd v_{f} \nonumber \\
&\hspace{6pt} \times\sum_{\kappa=-1}^{1}\abs{\bra{\nu_{E_{f}v_{f}}}\!\braket{ \psi_{E_{f}\lambda_{f}}|d_{\kappa}| \psi_{n_{i}} }\!\ket{\nu_{n_{i}v_{i}}}}^{2}.
\end{align}
Here, $\ket{\psi_{E_{f}\lambda_{f}}}$ denotes a continuum electronic state with energy $E_{f}=E_{i}+E_{\gamma}$ and pseudo--angular momentum $\lambda_{f}$, 
$\ket{\nu_{E_{f}v_{f}}}$ denotes a continuum vibrational level of the electronic state $\ket{\psi_{E_{f}\lambda_{f}}}$, 
$\ket{\nu_{n_{i}v_{i}}}$ denotes a bound vibrational level of the bound electronic state $\ket{\psi_{n_{i}}}$,
$c$ is the speed of light $\approx 137$, $\sigma_{\mathrm{T}} = 8\pi r_{0}/3 \approx 6.652 \times 10^{-25}$~cm$^{2}$ is the Thomson cross section, and $\braket{\Psi_{E_{f}v_{f}}|d_{\kappa}|\Psi_{n_{i}v_{i}}}$ is a dipole matrix element.
The $\kappa=0$ component of the dipole matrix element corresponds to photon polarization parallel to the internuclear axis and the $\kappa=\pm1$ components correspond to photon polarization perpendicular to the internuclear axis.
The final vibrational levels $\ket{\nu_{E_{f}v_{f}}}$ are assumed to be degenerate and can be integrated over using closure.
The photoionization cross section resolved in the initial vibrational level as a function of incident photon energy is then
\begin{equation}
\sigma_{n_{i}v_{i}}^{(\mathrm{PI})}(E_{\gamma}) = \braket{ \nu_{n_{i}v_{i}}|\sigma_{n_{i}}^{(\mathrm{PI})}(E_{\gamma};R)| \nu_{n_{i}v_{i}}}, \label{eq:vib_ion}
\end{equation}
which requires evaluating the electronic part of the photoionization cross section over a range of internuclear separations.
The electronic part of the photoionization cross section is defined as
\begin{equation}
\sigma_{n_{i}}^{(\mathrm{PI})}(E_{\gamma};R) = \sigma_{\mathrm{T}}\frac{\pi c^{3}E_{\gamma}}{2} \sum_{\lambda_{f}}\sum_{\kappa=-1}^{1}\abs{\braket{\psi_{E_{f}\lambda_{f}}|d_{\kappa}| \psi_{n_{i}}}}^{2}. \label{eq:FN_ion}
\end{equation}
Explicit forms of the dipole matrix elements are given in appendix~\ref{App:DipME}.

The rate coefficient for photoionization from the initial state $\ket{\Psi_{n_{i}v_{i}}}$ (per unit time) is
\begin{equation}
\mathcal{R}^{(\mathrm{PI})}_{n_{i}v_{i}} (T_{\gamma}) = \int_{E_{\mathrm{th}}}^{\infty} \!\rmd E_{\gamma}\, G_{\gamma}(E_{\gamma},T_{\gamma}) c \sigma_{n_{i}v_{i}}^{(\mathrm{PI})}(E_{\gamma}),
\end{equation}
where 
\begin{equation}
G_{\gamma}(E_{\gamma},T_{\gamma}) = \frac{8\pi E_{\gamma}^{2}}{h^{3}c^{3}} \frac{1}{\rme^{E_{\gamma}/k_{\mathrm{B}}T_{\gamma}}-1} \label{eq:spectral_dist}
\end{equation}
is the number of photons per unit volume per spectral unit $E_{\gamma}$ in the interval between $E_{\gamma}$ and $E_{\gamma}+\rmd E_{\gamma}$, $E_{\mathrm{th}}$ is the ionization threshold energy, $h$ is Planck's constant, $T_{\gamma}$ is the photon temperature, $k_{\mathrm{B}}$ is the Boltzmann constant, and all energies are in eV. 
The LTE photoionization rate coefficient is found by taking the thermal average of the vibrationally resolved rate coefficients,
\begin{equation}
\mathcal{R}_{\mathrm{LTE}}^{(\mathrm{PI})}(T_{\gamma}) =  \sum_{v_{i}=0}  \frac{g_{n_{i}v_{i}}(\mathrm{H}_{2}^{+})\,\rme^{-\varepsilon_{n_{i}v_{i}} /k_{\mathrm{B}}T_{\gamma}}}{\mathcal{Z}_{n_{i}}(T_{\gamma})} \,\mathcal{R}^{(\mathrm{PI})}_{n_{i}v_{i}} (T_{\gamma}),
\end{equation}
where $\varepsilon_{n_{i}v_{i}}$ is the energy of the state $\ket{\Psi_{n_{i}v_{i}}}$, $g_{n_{i}v_{i}}(\mathrm{H}_{2}^{+})$ are statistic weights~\citep{Stancil1994}, and 
\begin{equation}
\mathcal{Z}_{n_{i}}(T_{\gamma}) = \sum_{v_{i}=0}  g_{n_{i}v_{i}}(\mathrm{H}_{2}^{+})\,\rme^{-\varepsilon_{n_{i}v_{i}}/k_{\mathrm{B}}T_{\gamma}}
\end{equation}
is the partition function for the ground electronic state.
LTE photoionization cross sections can also be calculated:
\begin{align}
\sigma_{n_{i}}^{(\text{LTE--PI})}(E_{\gamma},T_{\mathrm{gas}}) &= \sum_{v_{i}=0}  \frac{g_{n_{i}v_{i}}(\mathrm{H}_{2}^{+})\,\rme^{-\varepsilon_{n_{i}v_{i}} /k_{\mathrm{B}}T_{\mathrm{gas}}}}{\mathcal{Z}_{n_{i}}(T_{\mathrm{gas}})} \nonumber \\
&\hspace{70pt} \times\sigma^{(\mathrm{PI})}_{n_{i}v_{i}} (E_{\gamma}),
\end{align}
where $T_{\mathrm{gas}}$ is the temperature of the H$_{2}^{+}$ gas.

\section{Results \& Discussion}
\label{sec:results}
In this section we present photoionization cross sections and rate coefficients for the ground and excited vibrational levels of H$_{2}^{+}$, HD$^{+}$, and D$_{2}^{+}$ within the ground electronic, 1s$\sigma_{\mathrm{g}}$.
Benchmark fixed--nuclei photoionization cross sections are presented and used to verify our results against previous calculations.
The model presented by \citet{Cohen1966} for the total cross section of H$_{2}^{+}$ is tested.
All cross sections are given in units of cm$^{2}$ and presented as functions of the incident photon energy in eV.
To obtain fixed--nuclei cross sections that represent an average over nuclear motion, calculations should be performed at the mean internuclear separation of the ground state, $R=2.06$ $a_{0}$. 
However, to compare with previous theoretical results we performed calculations at an internuclear separation of $R=2$ $a_{0}$.
Calculations are performed in both the length and velocity gauges and are found to produce identical results for all cross sections and incident photon energies considered. For clarity, only the velocity gauge results are presented.
Cross sections for photoionization from all bound vibrational states of H$_{2}^{+}$, HD$^{+}$, and D$_{2}^{+}$ are presented and isotopologue effects are investigated.
Energy levels for all bound vibrational states for H$_{2}^{+}$ and its deuterated isotopologues are presented in Table~\ref{tab:vib_en}.
Data files for our set of benchmark fixed--nuclei, vibrationally--resolved photoionization cross sections, and photoionization rate coefficients and fitting parameters for H$_{2}^{+}$ and its deuterated isotopologues are available on Zenodo under an open-source 
Creative Commons Attribution license: 
\dataset[doi:10.5281/zenodo.8304060]{https://doi.org/10.5281/zenodo.8304060}.

\begin{deluxetable*}{lDDD}
\tablenum{1}
\tablecaption{Total Bound vibrational energy levels $\varepsilon_{nv}$ for H$_{2}^{+}$ and its deuterated isotopologues. Energies are given in eV.}
\label{tab:vib_en}
\tablehead{
\colhead{} &  \multicolumn{6}{c}{Energy (eV)}  \\
\cline{2-7}
\colhead{$v~$} & \twocolhead{H$_{2}^{+}$} & \twocolhead{HD$^{+}$} &  \twocolhead{D$_{2}^{+}$}
}
\decimals
\startdata
0  & -16.2560 & -16.2749	 	& -16.2974	\\
1  & -15.9842 & -16.0376	 	& -16.1018 	\\
2  & -15.7282 & -15.8123		& -15.9143	\\
3  & -15.4875 & -15.5985	 	& -15.7345	\\
4  & -15.2616 & -15.3960	 	& -15.5624	\\
5  & -15.0501 & -15.2046	 	& -15.3978	\\
6  & -14.8529 & -15.0238	 	& -15.2406	\\
7  & -14.6696 & -14.8538	 	& -15.0905	\\
8  & -14.5001 & -14.6941	 	& -14.9476	\\
9  & -14.3444 & -14.5449	 	& -14.8116	\\
10 & -14.2027 & -14.4060	 	& -14.6827	\\
11 & -14.0748 & -14.2775	 	& -14.5607	\\
12 & -13.9613 & -14.1595	 	& -14.4456	\\
13 & -13.8623 & -14.0519	 	& -14.3374	\\
14 & -13.7785 & -13.9551	 	& -14.2361	\\
15 & -13.7104 & -13.8693	 	& -14.1418	\\
16 & -13.6588 & -13.7948	 	& -14.0546	\\
17 & -13.6247 & -13.7320	 	& -13.9745	\\
18 & -13.6086 & -13.6813	 	& -13.9017	\\
19 & -13.6058 & -13.6435	 	& -13.8363	\\
20 & .		  & -13.6189	 	& -13.7786	\\
21 & .		  & -13.6078	 	& -13.7287	\\
22 & .		  & -13.6057	 	& -13.6870	\\
23 & .		  & .	 			 	& -13.6538	\\
24 & .		  & .	 			 	& -13.6293	\\
25 & .		  & .	 				& -13.6139	\\
26 & .		  & .	 			 	& -13.6072	\\
27 & .		  & .	  			 	& -13.6057	\\
\enddata
\end{deluxetable*}

\subsection{Fixed--Nuclei Photoionization}
\label{sec:FN-ion}
\begin{figure}[!htb]
	\centering
	\includegraphics[width=0.92\linewidth]{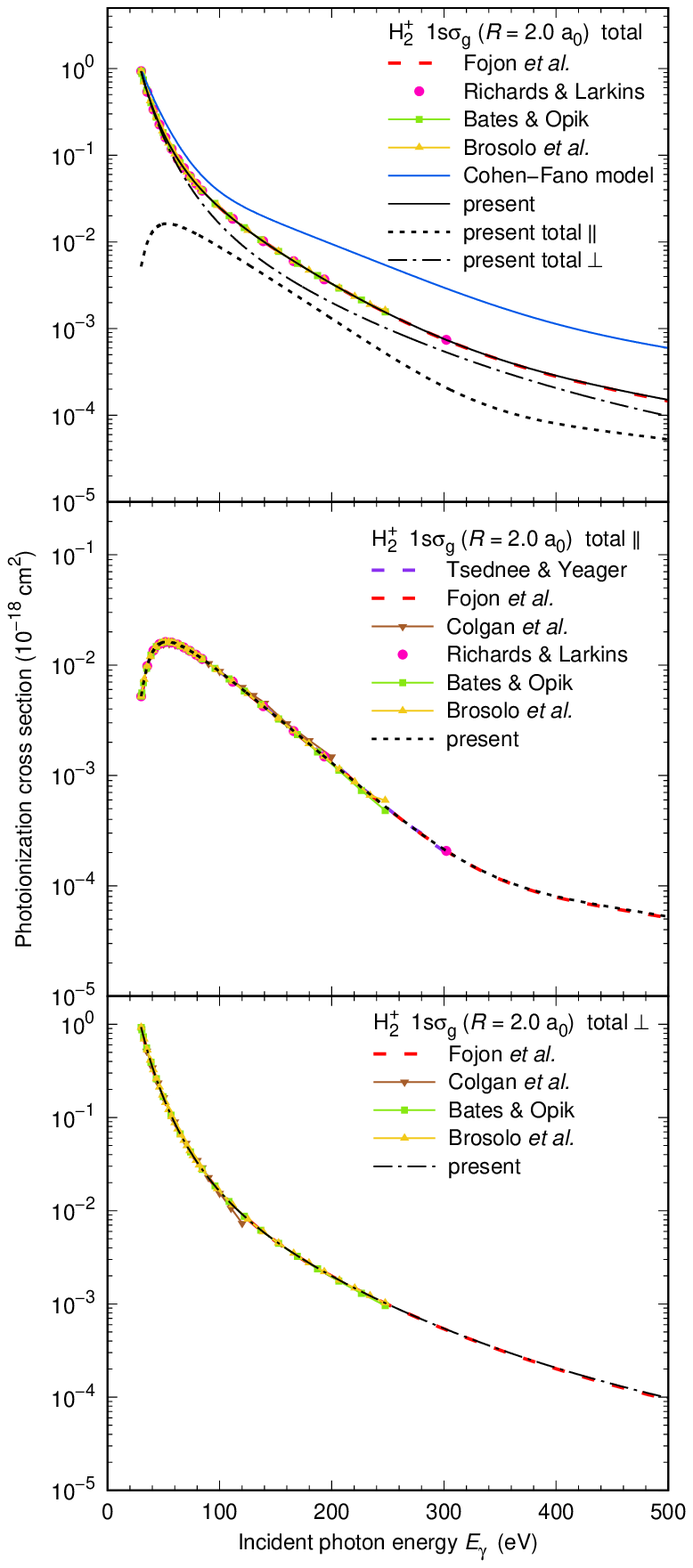}
	\caption{Total (solid black line), parallel (dotted black line), and perpendicular (dot--dashed black line) photoionization cross sections for the ground electronic state of H$_{2}^{+}$. We compare with the results of \citet{Fojon2006}, \citet{Richards1986}, \citet{Bates1968}, \citet{Brosolo1994}, \citet{Colgan2003}, \citet{Tsednee2018} and the \citet{Cohen1966} model. The Cohen--Fano model cross section has been shifted and rescaled to match ours at threshold.} \label{fig:ion_1sSg_total}
\end{figure}

The total photoionization cross section for the 1s$\sigma_{\mathrm{g}}$ state of H$_{2}^{+}$ is given by the solid line in Figure~\ref{fig:ion_1sSg_total}.
This cross section is the sum of the photoionization cross sections for the molecule oriented with the internuclear axis parallel (dotted lines) and perpendicular (dot--dashed lines) to the photon polarization.
Going forward we will refer to these cross sections as the parallel and perpendicular photoionization cross sections.

The perpendicular photoionization cross section is larger than the parallel photoionization cross section at all photon energies considered, the largest difference occurring near the ionization threshold.
The parallel and perpendicular photoionization cross sections presented here contain contributions from the lowest five partial waves, i.e. $\lambda = 1,3,5,7,9$.
For the energy range considered here higher partial waves are not required as convergence is achieved by $\lambda=9$.
Our cross sections are compared with those of \citet{Fojon2006}, \citet{Richards1986}, \citet{Bates1968}, \citet{Brosolo1994}, \citet{Colgan2003}, and \citet{Tsednee2018}. 
Excellent agreement is found in all cases.
In addition, though not presented, perfect agreement is found between our cross section and those of \citet{Bates1968} for photoionization of the 2s$\sigma_{\mathrm{g}}$ state of H$_{2}^{+}$.

In Figure~\ref{fig:ion_1sSg_total} we also compare our total photoionization cross section with the model, Equation~(\ref{eq:CF_model}), proposed by \citet{Cohen1966}. 
The ground state photoionization cross section for He$^{+}$ was used as the atomic cross section in the Cohen--Fano model.
After shifting the model photoionization cross section to the correct threshold and rescaling it to match our photoionization cross section at the ionization threshold, we find that the Cohen--Fano model cross section has similar qualitative behavior at low energies but at high energies it does not decay as quickly as our photoionization cross section.
This leads to significant differences in magnitude at higher incident photon energies, which is in agreement with the observations of \citet{Fojon2006}.
The Cohen--Fano model makes three assumptions: (i) ionization is effectively a one--electron process, (ii) the initial state wave function is well described within the Linear Combination of Atomic Orbitals approximation, 
and (iii) the ionized electron is well described by a plane wave or single--center spherical wave.
In the case of H$_{2}^{+}$ the first of these assumptions is valid, the validity of the second assumption depends on the size of the basis of atomic orbitals, and the third assumption is not satisfied. Hence, it is not surprising that our photoionzation cross section differs from the Cohen--Fano model photoionization cross section.

\subsection{Vibrationally--Resolved Photoionization}
\label{sec:vib_ion}
\begin{figure}[!htb]
	\centering
	\includegraphics[width=0.99\linewidth]{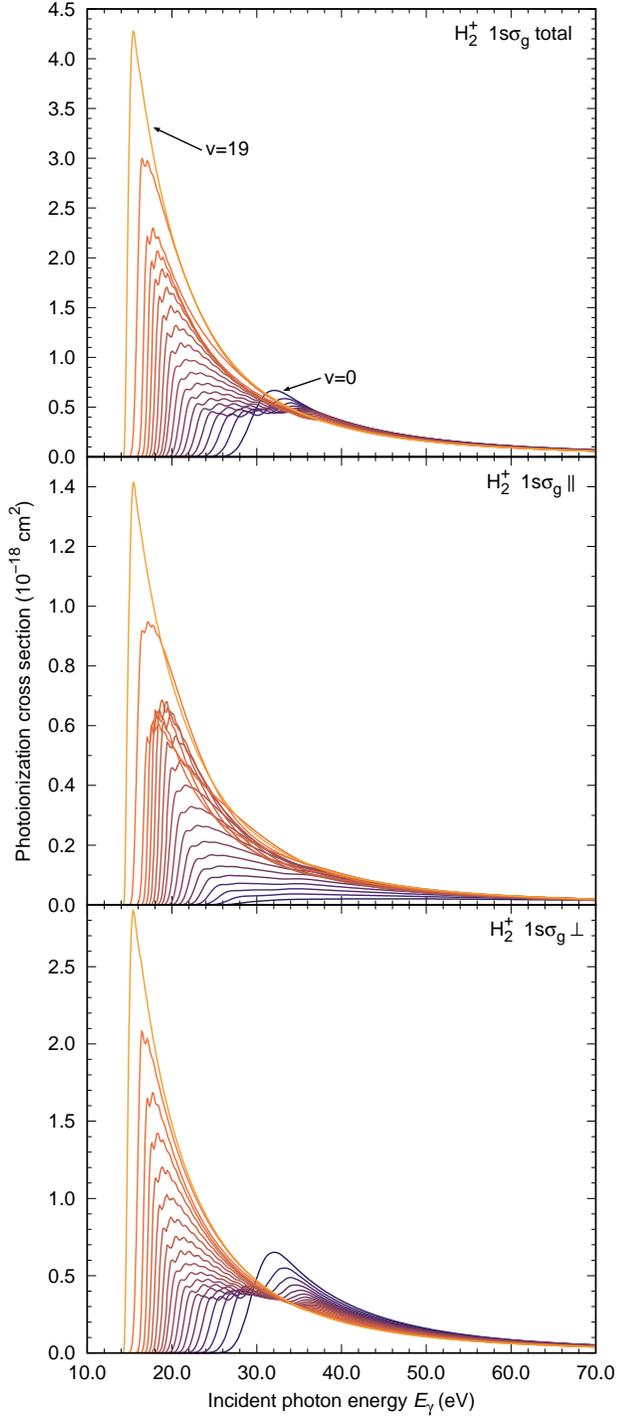}
	\caption{Total, parallel, and perpendicular photoionization cross sections for all bound vibrational levels of the 1s$\sigma_{\mathrm{g}}$ state of H$_{2}^{+}$.} \label{fig:vib_all}
\end{figure}

Total, parallel, and perpendicular photoionization cross sections for all bound vibrational levels of the 1s$\sigma_{\mathrm{g}}$ state are shown in Figure~\ref{fig:vib_all}.
The total photoionization cross sections are typically dominated by perpendicular contributions.
For $v\geq 4$, the cross sections near their respective thresholds increase significantly in magnitude as the vibrational level $v$ increases.
The increase in the magnitude of the cross section as $v$ increases is also observed in electron scattering and photodissociation of H$_{2}^{+}$~\citep{Zammit2014,Scarlett2017,Zammit2017}, and electron scattering from H$_{2}$~\citep{Scarlett2021}.
However, for $v\leq 3$ the near threshold cross section decreases as $v$ increases.
This behaviour in the total cross section follows from the perpendicular photoionization cross sections.
Oscillations in the cross sections just above the ionization threshold can also be seen. 
We believe these oscillations are physical as the only approximation used in these calculations is the Born--Oppenheimer approximation. 
Similar behaviour is also observed in electron scattering cross sections \citep{Scarlett2021}.
The ionization thresholds also shifts to lower energies as the initial vibrational level increases.

The origin of the oscillations present in the near threshold cross sections for photoionization from excited vibrational levels is investigated further in appendix~\ref{App:osc}.

\subsection{Isotopologue Effects}
\label{sec:isotopologue}

\begin{figure}[!htb]
	\centering
	\includegraphics[width=0.99\linewidth]{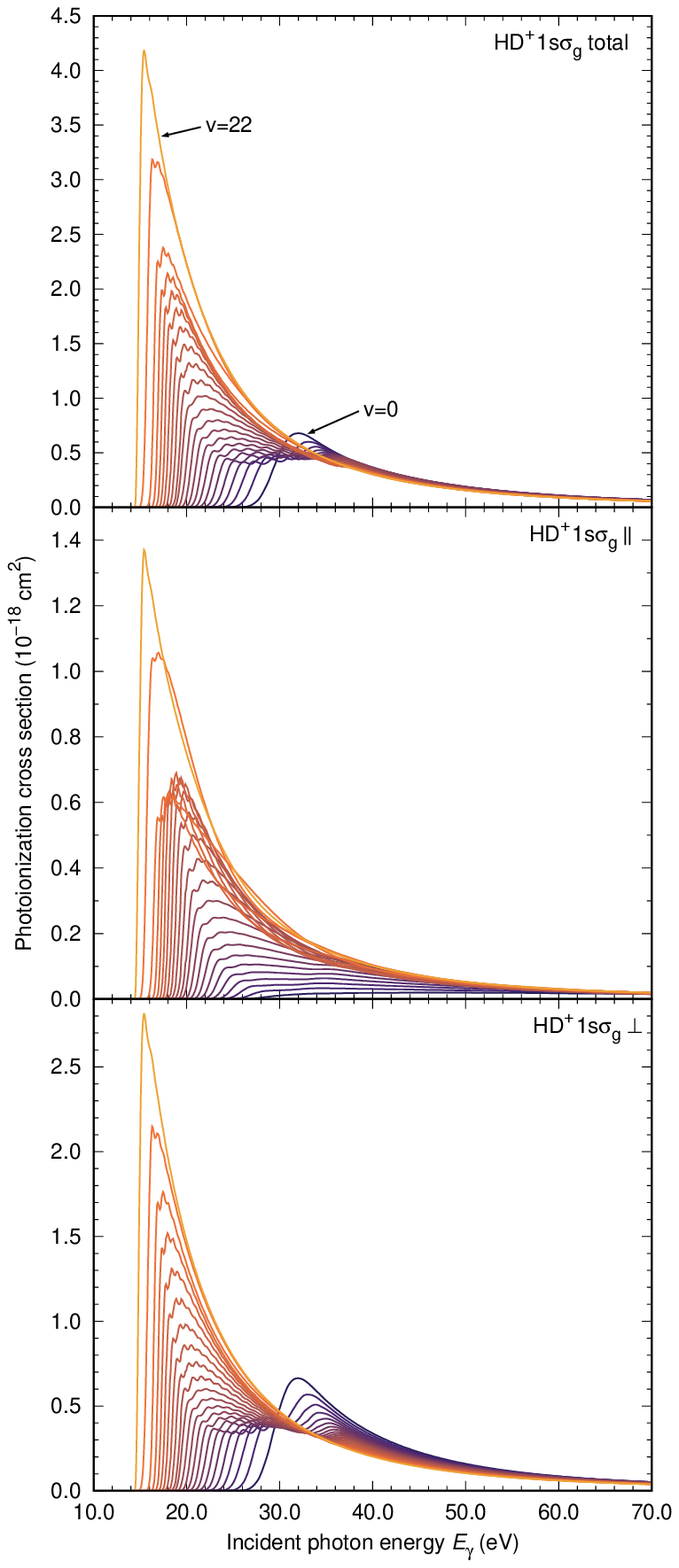}
	\caption{Same as Figure~\ref{fig:vib_all} but for HD$^{+}$.} \label{fig:HD_vib_all}
\end{figure}

\begin{figure}[!htb]
	\centering
	\includegraphics[width=0.99\linewidth]{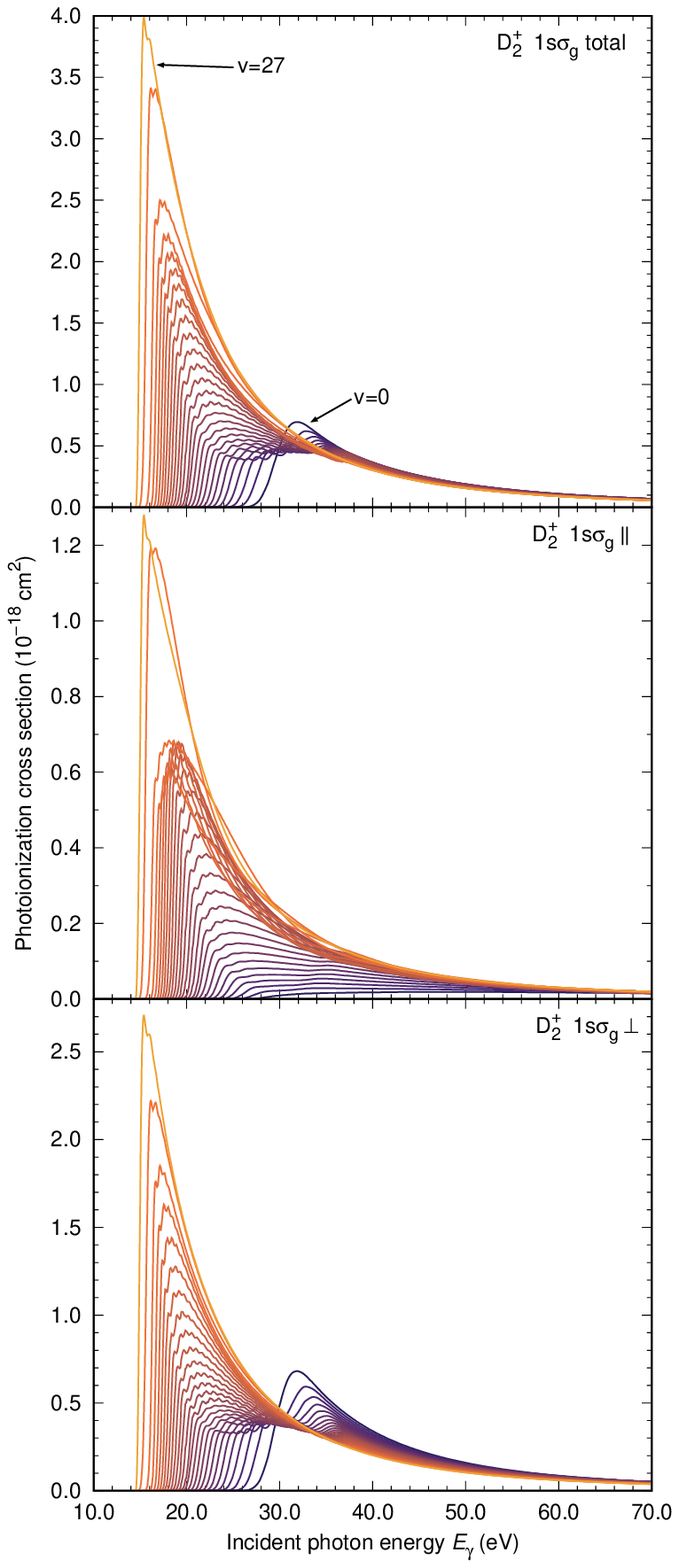}
	\caption{Same as Figure~\ref{fig:vib_all} but for D$_{2}^{+}$.} \label{fig:D2_vib_all}
\end{figure}

In Figures~\ref{fig:HD_vib_all} and \ref{fig:D2_vib_all} we present photoionization cross sections for all bound vibrational levels of the ground electronic state of HD$^{+}$, and D$_{2}^{+}$ respectively.
These cross section exhibit the same behavior as the H$_{2}^{+}$ cross sections, i.e. the cross section increasing in magnitude near the ionization threshold with increasing vibrational level and oscillations in the cross sections occurring near the ionization threshold for photoionization from excited vibrational levels.

\begin{figure}[!htb]
	\centering
	\includegraphics[width=0.99\linewidth]{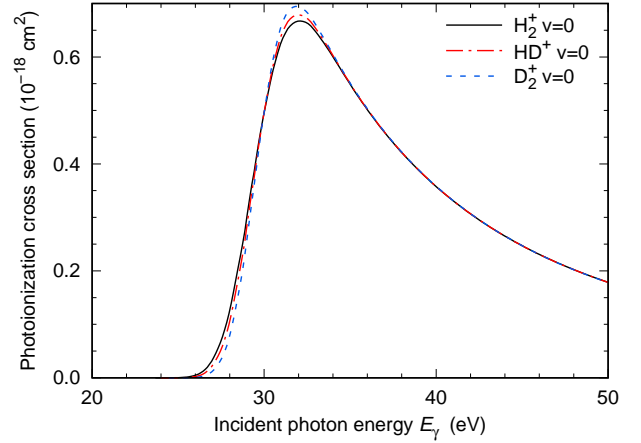}
	\caption{Total photoionization cross sections for the ground state of H$_{2}^{+}$ and its deuterated isotopologues.} \label{fig:vib_H2D2T2}
\end{figure}

The cross sections for photoionization from the ground vibrational level of H$_{2}^{+}$ and its deuterated isotopologues are compared in Figure~\ref{fig:vib_H2D2T2}.
There are only minor differences between the cross sections for each isotopologue, with the maximum value of the cross section increasing slightly and the ionization threshold shifting to a higher energy as the nuclear reduced mass of the isotopologue increases.
This is due to the energy of the ground vibrational level decreasing as the nuclear reduced mass increases, which results in the ionization threshold shifting to a higher energy.
The increase in the maximum value of the photoionization cross section for heavier isotoplogues is a result of the vibrational wave function being more densely concentrated in the region where the electronic part of the photoionization cross section is larger.
At higher incident photon energies the photoionization cross sections for the different isotopologues are identical.

\begin{figure}[!htb]
	\centering
	\includegraphics[width=0.99\linewidth]{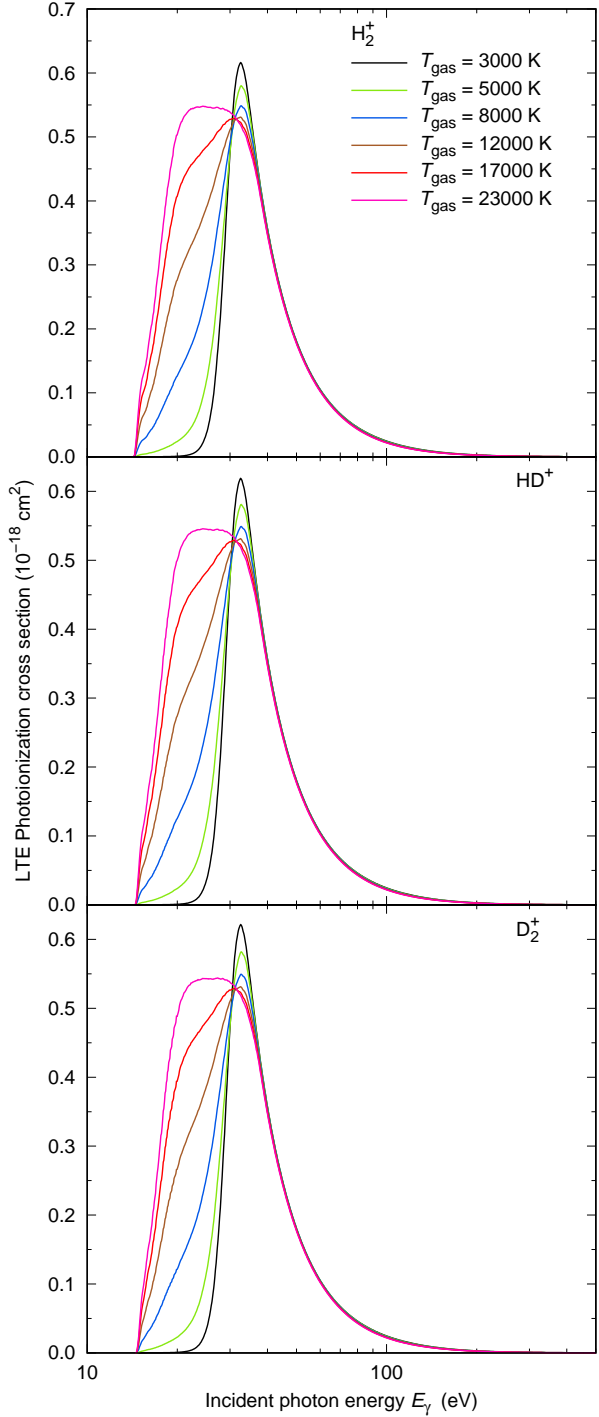}
	\caption{LTE photoionization cross sections for H$_{2}^{+}$ and its deuterated isotopologues at gas temperatures of $3\,000$, $5\,000$, $8\,000$, $12\,000$, $17\,000$, and $23\,000$ K.} \label{fig:LTE_xsec}
\end{figure}

In Figure~\ref{fig:LTE_xsec}, LTE photoionization cross sections for H$_{2}^{+}$ and its deuterated isotopologues are presented over a range of gas temperatures.
These cross sections can be integrated with arbitrary spectral energy density distributions to obtain photoionization rate coefficients. 

\subsection{Photoionization Rate Coefficients}
\label{sec:rates}
\begin{figure}[!htb]
	\centering
	\includegraphics[width=0.99\linewidth]{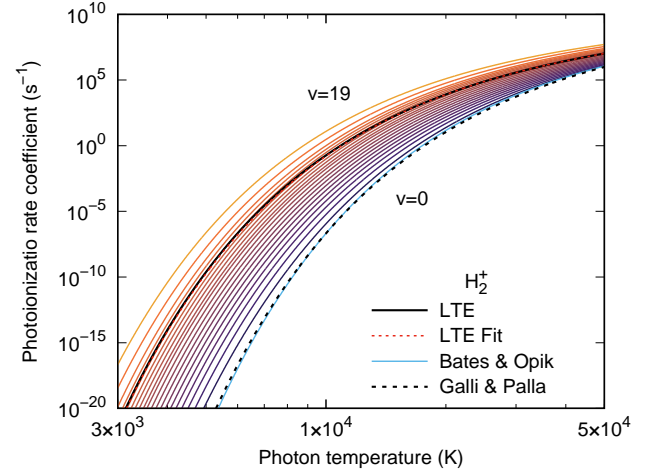}
	\caption{Vibrationally--resolved and LTE photoionization rate coefficients for H$_{2}^{+}$ as a function of radiation temperature. The analytic fit of \citet{Galli1998} and the rate coefficient we have calculated using the cross sections of \citet{Bates1968} are presented for comparison. The analytic fit function in Equation~(\ref{eq:fit}) is presented for the LTE photoionization rate coefficients.} \label{fig:H2rates}
\end{figure}

\begin{figure}[!htb]
	\centering
	\includegraphics[width=0.99\linewidth]{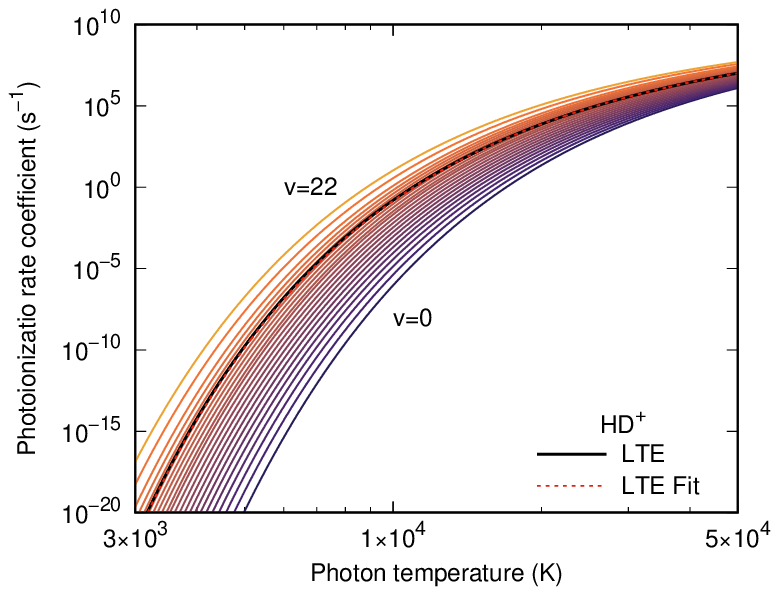}
	\caption{Same as Figure~\ref{fig:H2rates} but for HD$^{+}$.} \label{fig:HDrates}
\end{figure}

\begin{figure}[!htb]
	\centering
	\includegraphics[width=0.99\linewidth]{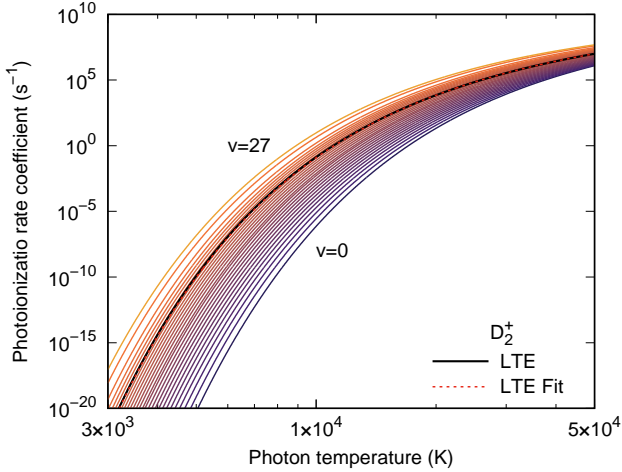}
	\caption{Same as Figure~\ref{fig:H2rates} but for D$_{2}^{+}$.} \label{fig:D2rates}
\end{figure}

Vibrationally--resolved and LTE photoionization rate coefficients are given Figures~\ref{fig:H2rates}--\ref{fig:D2rates}, with analytic fits included for the LTE rate coefficients.
As discussed in Section~\ref{sec:isotopologue}, there are practically no isotopologue effects present in the photoionization cross sections which results in the LTE photoionization rate coefficients for each isotopologue being very similar.

\begin{figure}[!htb]
	\centering
	\includegraphics[width=0.99\linewidth]{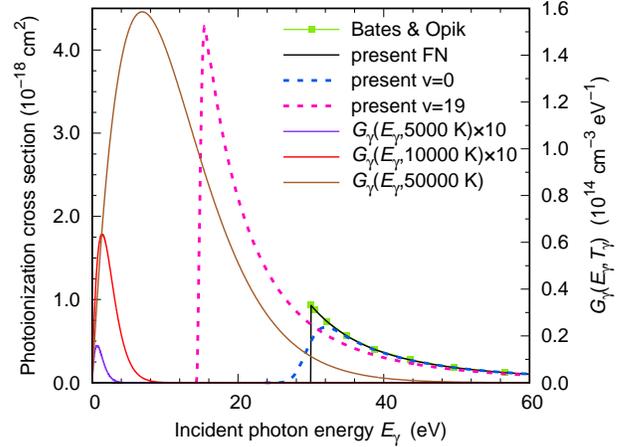}
	\caption{Present fixed--nuclei $R=2 ~a_{0}$, and vibrationally--resolved $v=0$ and $v=19$ photoionization cross sections superimposed with the spectral energy density distribution given by Equation~(\ref{eq:spectral_dist}). The fixed--nuclei photoionization cross sections of \citet{Bates1968} are presented for comparison. The spectral energy density distributions at $5\,000$ K and $10\,000$ K have been scaled by a factor of 10 for clarity.} \label{fig:test_Bates}
\end{figure}

The analytic model given by \citet{Galli1998} was determined by fitting to a photoionization rate coefficient calculated using the fixed--nuclei cross sections of \citet{Bates1968}.
In Figure~\ref{fig:H2rates} we verify our rate coefficients by calculating a photoionzation rate coefficient using the results of \citet{Bates1968} and confirming that it agrees with the analytic fit of \citet{Galli1998}.
The cross sections of \citet{Bates1968} are calculated within the fixed--nuclei approximation, which would approximate the $v=0$ photoionization cross section. 
Hence, photoionization rate coefficients calculated using the \citet{Bates1968} cross sections are similar to our $v=0$ rate coefficients with the difference being smallest at higher temperatures.
Photoionization rate coefficients for excited vibrational levels, in Figure~\ref{fig:H2rates}, are much larger than the rate coefficient for the ground vibrational level.
This results in the LTE photoionization rate coefficient being several orders of magnitude larger than the currently used values of~\citet{Galli1998}.
The significant variance in the magnitude of the vibrationally--resolved rate coefficients is due to the ionization threshold for excited vibrational levels shifting to lower photon energies as seen in Figure~\ref{fig:test_Bates}.
This leads to a larger overlap between the cross section and the spectral energy density distribution, Equation~(\ref{eq:spectral_dist}), resulting in a larger photoionization rate coefficient.

We use the analytic fit function of \citet{Galli1998},
\begin{equation}
\mathcal{R}^{(\mathrm{PI})}_{n_{i}v_{i}} (T_{\gamma}) = A \,T_{\gamma}^{B} \,\rme^{C/T_{\gamma}} \label{eq:fit}
\end{equation}
and determine fitting parameters $A$, $B$, and $C$ for the vibrationally--resolved and LTE photoionization rate coefficients for H$_{2}^{+}$ and its deuterated isotopologues. 
Fitting parameters for the LTE photoionization rate coefficient for H$_{2}^{+}$, HD$^{+}$, and D$_{2}^{+}$ are presented in Table~\ref{tab:params}, and were determined using standard non--linear least--squares curving fitting.
Fitting parameters for the vibrationally--resolved rate coefficients are available on Zenodo under an open-source 
Creative Commons Attribution license: 
\dataset[doi:10.5281/zenodo.8304060]{https://doi.org/10.5281/zenodo.8304060}.

\begin{deluxetable}{cDDC}[h]
\tablenum{2}
\tablecaption{Fitting parameters in Equation~\ref{eq:fit} for the LTE photoionization rate coefficient for H$_{2}^{+}$ and its deuterated isotopologues.}
\label{tab:params}
\tablehead{
\colhead{isotopologue} & \twocolhead{A (s$^{-1}$ K$^{-B}$)} & \twocolhead{B} & \colhead{C (K)}
}
\decimals
\startdata
H$_{2}^{+}$  & 113.13 & 1.42 & -196236 \\
HD$^{+}$     & 134.22 & 1.41 & -197978 \\
D$_{2}^{+}$  & 150.67 & 1.40 & -199736 \\
\enddata
\end{deluxetable}

\section{Conclusion}
\label{sec:conclusion}
We have calculated cross sections and rate coefficients for photoionization from all bound vibrational states of H$_{2}^{+}$, HD$^{+}$, and D$_{2}^{+}$ and presented a benchmark set of H$_{2}^{+}$ fixed--nuclei photoionization cross sections.
Vibrationally--resolved and LTE photoionization rate coefficients are calculated for H$_{2}^{+}$ and its deuterated isotopologues.
The present vibrationally--resolved photoionization rate coefficients are found to be several orders of magnitudes larger than the currently accepted values in the literature.
Fitting parameters for an analytic model of the vibrationally--resolved and LTE photoionization rate coefficients have been provided for modeling the effects of photoionization of H$_{2}^{+}$ and its isotopologues in astrophysical plasmas.

Our benchmark results are found to be in excellent agreement with available theoretical results in all cases.
It was found that the Cohen--Fano model photoionization cross section can describe the overall shape of the H$_{2}^{+}$ photoionization cross section, but does not reproduce the correct magnitude.
Near threshold cross sections for photoionization from vibrationally excited levels of H$_{2}^{+}$ and its isotopologues were found to contain oscillations.
These oscillations were shown to be due to nodes in the vibrational wave functions. 
The magnitude of the vibrationally--resolved photoionization cross sections increases as the initial vibrational level increases, while the ionization threshold shifts to lower energies.
No significant isotopologue effects were found in the cross sections.
Data files for our set of benchmark cross sections, rate coefficients, and fitting parameters for H$_{2}^{+}$, HD$^{+}$, and D$_{2}^{+}$ are available on Zenodo under an open-source 
Creative Commons Attribution license: 
\dataset[doi:10.5281/zenodo.8304060]{https://doi.org/10.5281/zenodo.8304060}.
Using the data supplied with this manuscript the reader can construct thermally--averaged photoionization cross sections, which can then be integrated with an arbitrary spectral energy density distribution to obtain photoionization rate coefficients.

\begin{acknowledgments}
This work was supported by the Government of Western Australia Defense Science Centre Research Higher Degree Student Grant, and by the United States Air Force Office of Scientific Research, grant number FA2386-19-1-4044.
HPC resources were provided by the Pawsey Supercomputing Centre with funding from the Australian Government and the Government of Western Australia. 
A.S. acknowledges the contribution of an Australian Government Research Training Program Scholarship. 
M.C.Z would like to specifically acknowledge the support of the Los Alamos National Laboratory (LANL) ASC PEM Atomic Physics Project. LANL is operated by Triad National Security, LLC, for the National Nuclear Security Administration of the U.S. Department of Energy under Contract No. 89233218NCA000001. 
\end{acknowledgments}

\pagebreak

\appendix

\section{Molecular Target Structure}
\label{App:structure}
We start by applying the Born--Oppenheimer approximation which assumes that the electronic and nuclear parts of the target wave function are separable, Equation~(\ref{eq:BO}). 
The unseparated electronic Hamiltonian of H$_{2}^{+}$ and its isotopologues in prolate spheroidal coordinates is given by
\begin{equation}
H = \frac{-1/2}{\left(\rho+\tfrac{R}{2}\right)^{2} - \left(\tfrac{R}{2}\eta\right)^{2}} \left\{ \frac{\partial}{\partial\rho}\left[\rho(\rho+R)\frac{\partial}{\partial\rho} \right] + \frac{\partial}{\partial\eta}\left[(1-\eta^{2})\frac{\partial}{\partial\eta} \right] + \left[ \frac{R^{2}/4}{\rho(\rho+R)} + \frac{1}{1-\eta^{2}}\right] \frac{\partial^{2}}{\partial\phi^{2}} + 2(2\rho+R) \right\}, \label{eq:unsep_hamiltonian}
\end{equation}
where the quasi--radial and quasi--angular coordinates are
\begin{equation}
\qquad \rho =\frac{r_{1}+r_{1}}{2}-\frac{R}{2}, \qquad\qquad \eta = \frac{r_{1}-r_{2}}{R}.
\end{equation}
Here $r_{1}$ and $r_{2}$ are the coordinates of the electron relative to the two nuclei, $R$ is the internuclear separation, $\phi$ is the usual azimuthal coordinate.
Bound electronic states are obtained by diagonalizing the unseparated electronic Hamiltonian, Equation~(\ref{eq:unsep_hamiltonian}), in a Sturmian basis $\{ \varphi_{k\ell m }\}$, and are represented by
\begin{equation} \label{eq:bound_st_expansion}
\psi_{n}^{m}(\rho,\eta,\phi;R) = \sum_{k=1}^{k_{\mathrm{max}}} \sum_{\ell=\abs{m}}^{\ell_{\mathrm{max}}} C^{(n)}_{k\ell} \varphi_{k\ell m}(\rho,\eta,\phi),
\end{equation}
where 
\begin{equation}
\varphi_{k\ell m}(\rho,\eta,\phi) = f_{k}^{m}(\rho) Y_{\ell}^{m}(\eta,\phi), \label{eq:basis}
\end{equation}
\begin{align}
f_{k}^{m}(\rho) = \sqrt{2\alpha_{m}\frac{(k-1)!}{(k+m-1)!}}(2\alpha_{m}\rho)^{m/2} \rme^{-\alpha_{m}\rho} L_{k-1}^{m}(2\alpha_{m}\rho).
\end{align}
Here, $L_{k-1}^{m}$ are the associated Laguerre polynomials, $Y_{\ell}^{m}$ are spherical harmonics, $\alpha_{m}$ is an exponential fall--off parameter, and $k$ is an index.
These basis functions are orthogonal with respect to their coordinates: 
\begin{align}
\int_{0}^{\infty} \rmd\rho \, f_{k'}^{m}(\rho)f_{k}^{m}(\rho) = \delta_{k'k}, \\
\int_{0}^{2\pi}\!\rmd\phi \int_{-1}^{1}\!\rmd\eta\, Y_{\ell'}^{m'*}(\eta,\phi)Y_{\ell}^{m}(\eta,\phi) = \delta_{\ell'\ell}\delta_{m'm},
\end{align}
but not the volume element in prolate spheroidal coordinates. 
The overlap between the basis functions, Equation~(\ref{eq:basis}), with the volume element
\begin{equation}
\rmd V = \left[\left(\rho+\tfrac{R}{2} \right)^{2} - \left(\tfrac{R}{2}\eta \right)^{2} \right]\rmd\rho\rmd\eta \rmd \phi,
\end{equation}
is given by~\citep{SavagePhD}
\begin{align}\label{eqA:ovrlap}
\braket{k'\ell'm'|k\ell m} &= \int \rmd V \,\varphi^{*}_{k'\ell'm'}(\rho,\eta,\phi)\varphi_{k\ell m}(\rho,\eta,\phi) \nonumber \\
&=  \delta_{\ell'\ell}\delta_{m'm}b_{k'k}^{m} - \delta_{k'k}\tfrac{R^{2}}{6}C_{\ell'020}^{\ell0}C_{\ell m 20}^{\ell'm'},
\end{align}
where 
\begin{equation}
b_{k'k}^{m} = \begin{cases}
(2\alpha_{m})^{-2}\sqrt{k(k+m)(k+1)(k+m+1)} & k'=k+2,  \\
-(2\alpha_{m})^{-2}\sqrt{k(k+m)} (4k+2m+2\alpha_{m}R) & k'=k+1,  \\
(2\alpha_{m})^{-2}(6k^{2}-6k+6km -3m+m^{2}+2) +\tfrac{2k-1+m}{2\alpha_{m}} + \tfrac{R^{2}}{6} & k'=k,  \\
-(2\alpha_{m})^{-2}\sqrt{(k-1)(k-1+m)}(4k-4+2m+2\alpha_{m}R) & k'=k-1,  \\
(2\alpha_{m})^{-2}\sqrt{(k-1)(k-1+m)(k-2)(k-2+m)} & k'=k-2, \\
0 &$otherwise$,
\end{cases}
\end{equation}
and $C_{\ell m \ell' m'}^{LM}$ is a Clebsh--Gordon Coefficient.
The expansion coefficients, $C^{(n)}_{k\ell}$ in Equation~(\ref{eq:bound_st_expansion}), and the bound state energy levels, $E$, are obtained by solving the generalized eigenvalue problem
\begin{equation}
\vec{H} \vec{C} = E \vec{B} \vec{C}. 
\end{equation}
Here $\vec{H}$ is the Hamiltonian matrix with matrix elements given by 
\begin{align}\label{eqA:Ham_me}
\braket{k'\ell' m'|H|k\ell m} = \delta_{\ell'\ell}\delta_{m'm}\Bigg[ h_{k'k}^{\ell m} - \frac{m^{2}R}{8}\int_{0}^{\infty}\!\rmd \rho\, \frac{f_{k'}^{m}(\rho)f_{k}^{m}(\rho)}{\rho+R}\Bigg],
\end{align}
where
\begin{equation}
h_{k'k}^{\ell m} = \begin{cases}
-\tfrac{1}{8}\sqrt{k(k+m)(k+1)(k+1+m)} & k'=k+2, \\
\tfrac{1}{4}\sqrt{k(k+m)}\left(\alpha_{m}R - \tfrac{4}{\alpha_{m}} \right) & k'=k+1, \\
\tfrac{1}{4}\bigg[k^{2}-k+km-\tfrac{m}{2} +1+4R+2\ell(\ell+1) + (2k-1+m)\left( \alpha_{m}R+ \tfrac{4}{\alpha_{m}} \right) \bigg] & k'=k, \\
\tfrac{1}{4}\sqrt{(k-1)(k-1+m)}\left(\alpha_{m}R - \tfrac{4}{\alpha_{m}} \right) & k'=k-1, \\
-\tfrac{1}{8}\sqrt{(k-1)(k-1+m)(k-2)(k-2+m)} & k'=k-2, \\
0 &$otherwise$,
\end{cases}
\end{equation}
and $\vec{B}$ is the overlap matrix with elements given by Equation~(\ref{eqA:ovrlap}), and $\vec{C}$ contains the expansion coefficients as its columns.\\

Electronic continuum states are calculated using the approach of~\citet{Singor2022}.
The non--relativistic electronic Schr\"odinger equation for the molecular hydrogen ion is separable into quasi--radial and quasi--angular parts in prolate spheroidal coordinates and is given by
\begin{align}
\Bigg\{ \frac{\rmd}{\rmd\rho} \left[ \rho(\rho+R)\frac{\rmd}{\rmd\rho}\right] - \frac{m^{2}R^{2}}{4\rho(\rho+R)} + 2(2\rho+R) +2E\rho(\rho+R) - A_{\lambda}^{\abs{m}}(E,R) \Bigg\}\Xi_{\lambda}^{\abs{m}}(\rho;R) &= 0, \label{eq:radial} \\
\Bigg\{ \frac{\rmd}{\rmd\eta} \left[ (1-\eta^{2})\frac{\rmd}{\rmd\eta}\right] - \frac{m^{2}}{1-\eta^{2}} +\tfrac{1}{2}ER^{2}(1-\eta^{2})  + A_{\lambda}^{\abs{m}}(E,R) \Bigg\}\Upsilon_{\lambda}^{m}(\eta,\phi;R) &= 0, \label{eq:angular}  
\end{align}
where $E$ is the energy eigenvalue, $A_{\lambda}^{\abs{m}}(E,R)$ is the separation constant, $m$ is the projection of the orbital angular momentum onto the internuclear axis, and $\lambda$ is a pseudo--angular momentum used to label states.
The pseudo--angular momentum $\lambda$ is related to the number of zeros $q$, of $\Upsilon_{\lambda}^{m}(\eta,\phi;R)$ in the interval $\eta \in(-1,1)$ by $\lambda=q+\abs{m}$ and corresponds to the orbital angular momentum in the combined nuclei limit, i.e. $\lambda \to \ell$ as $R\to 0$.
The quasi--angular wave functions are obtained by expanding the spheroidal harmonics in Equation~(\ref{eq:angular}) as a series of spherical harmonics of the same $m$. The solution to the quasi--radial equation, Equation~(\ref{eq:radial}), is started using a power series expansion and then propagated using an Adams--Moulton predictor--corrector algorithm. An asymptotic series is then used to normalize the wave function.
The electronic continuum states of a H$_{2}^{+}$ isotopologue can then be written as 
\begin{align}
\psi_{E\lambda}^{m}(\rho,\eta,\phi;R) &= \Xi_{\lambda}^{\abs{m}}(\rho;R)\Upsilon_{\lambda}^{m}(\eta,\phi;R), \\
&= \Xi_{\lambda}^{\abs{m}}(\rho;R) \sum_{\ell=\abs{m}}^{\ell_{\mathrm{max}}} \mathcal{C}_{\ell m}^{\lambda}(R)\, Y_{\ell}^{m}(\eta,\phi).
\end{align}

The non--relativistic Born--Oppenheimer Schr\"odinger equation for the vibrational wave functions is
\begin{equation}
\left[ -\frac{1}{2\mu}\frac{\rmd^{2}}{\rmd R^{2}} +\epsilon_{n}(R) - \varepsilon_{nv}\right] \nu_{nv}(R)=0, \label{eq:vib_SE}
\end{equation}
where $n$ specifies the electronic state, $\epsilon_{n}(R)$ is the potential energy curve for the electronic state $n$, $\mu$ is the nuclear reduced mass of the H$_{2}^{+}$ isotopologue, and the centrifugal term is neglected in the current non--rotationally resolved calculations. 
Within the Born--Oppenheimer approximation the only difference between the isotopologues is the value of $\mu$.
The values of the nuclear reduced mass for H$_{2}^{+}$, HD$^{+}$, and D$_{2}^{+}$ are given in Table~\ref{tab:mass}, along with the number of bound vibrational levels.
Bound vibrational wave functions are found by expanding $\nu_{nv}(R)$ as
\begin{equation}
\nu_{nv}(R) = \sum_{k=1}^{k_{\mathrm{max}}} \tilde{C}_{knv} \,\phi_{k}(R), \label{eq:vib_wf}
\end{equation}
where 
\begin{equation}
\phi_{k}(R) = \frac{\sqrt{\tilde{\alpha}}}{k} \,2\tilde{\alpha}R \,\rme^{-\tilde{\alpha}R} \,L^{1}_{k-1}(2\tilde{\alpha}R).
\end{equation}
Here $\tilde{\alpha}$ is an exponential fall--off parameter, and $L^{1}_{k-1}$ are the associated Laguerre polynomials.
The expansion coefficients $\tilde{C}_{knv}$ in Equation~(\ref{eq:vib_wf}) are obtained by diagonalizing the vibrational Hamiltonian in the basis $\{\phi_{k}\}$.

\begin{deluxetable}{cDC}[h]
\tablenum{3}
\tablecaption{Nuclear reduced masses in atomic units, and number of bound vibrational levels in the ground electronic state for H$_{2}^{+}$ and its deuterated isotopologues, where $m_{\mathrm{e}}$ is the electron mass.}
\label{tab:mass}
\tablehead{
\colhead{\multirow{2}{*}{isotopologue}} & \twocolhead{\multirow{2}{*}{$\mu$ ($m_{\mathrm{e}}$)}} & \colhead{number of bound} \\
& && \colhead{vibrational levels} 
}
\decimals
\startdata
H$_{2}^{+}$  & 918.076351 & 20 \\
HD$^{+}$     & 1223.899246 & 23 \\
D$_{2}^{+}$  & 1835.241507 & 28 \\
\enddata
\end{deluxetable}

\section{Dipole Matrix Elements}
\label{App:DipME}
Matrix elements of the dipole operator can be given in the length $(\mathsf{L})$ or velocity $(\mathsf{V})$ gauge,
\begin{equation}
\braket{\psi_{E\lambda}^{m'}|d_{\kappa}|\psi_{n}^{m}} = \braket{\psi_{E\lambda}^{m'}|\mathsf{L}_{\kappa}|\psi_{n}^{m}} = \frac{\braket{\psi_{E\lambda}^{m'}|\mathsf{V}_{\kappa}|\psi_{n}^{m}}}{E_{n}-E}.
\end{equation}
The length gauge dipole operator is
\begin{align}
\mathsf{L}_{\kappa} &= \mathcal{L}_{\kappa}(\rho) \sqrt{\tfrac{4\pi}{3}} \,Y_{1}^{\kappa}(\eta,\phi), \\
\mathcal{L}_{\kappa}(\rho) &= \begin{cases}
\rho +\tfrac{R}{2} & \kappa=0, \\
\sqrt{\rho(\rho+R)} & \kappa=\pm1,
\end{cases}
\end{align}
and the velocity gauge dipole operator is
\begin{align}
\mathsf{V}_{\kappa} &= \begin{cases}
\dfrac{1}{\rho(\rho+R) + \frac{R^{2}}{4}(1-\eta^{2})}\left[\rho(\rho+R)\eta\dfrac{\partial}{\partial\rho} + (1-\eta^{2})\left(\rho+\frac{R}{2}\right)\dfrac{\partial}{\partial\eta} \right] & \kappa=0, \\[10pt]
\dfrac{\mp\rme^{\pm\rmi\phi}}{\sqrt{2}}\dfrac{\sqrt{\rho(\rho+R)}}{\rho(\rho+R) + \frac{R^{2}}{4}(1-\eta^{2})}\left[\sqrt{1-\eta^{2}}\left(\rho+\frac{R}{2} \right)\dfrac{\partial}{\partial\rho} \pm \rmi \dfrac{R^{2}}{4} \dfrac{\sqrt{1-\eta^{2}}}{\rho(\rho+R)}\dfrac{\partial}{\partial\phi}  \right. \\
\hspace{140pt}-\left. \left(\eta\sqrt{1-\eta^{2}}\dfrac{\partial}{\partial\eta} \mp \dfrac{\rmi}{\sqrt{1-\eta^{2}}}\dfrac{\partial}{\partial\phi} \right)\right] & \kappa=\pm1.
\end{cases}
\end{align}
The components of the electronic length gauge dipole matrix elements are given by~\citep{Zammit2017}
\begin{align}
\braket{\psi_{E\lambda}^{m'}|\mathsf{L}_{\kappa}|\psi_{n}^{m}} = \sum_{\ell'}\sum_{k\ell} \mathcal{C}_{\ell'm'}^{\lambda} C^{(n)}_{k\ell} \int_{0}^{\infty}\rmd\rho\,\,  \Xi_{\lambda}^{\abs{m'}}(\rho)\, \mathcal{L}_{\kappa}(\rho)  f_{k}^{m}(\rho)J_{1,\kappa}^{(\ell'm',\ell m)}(\rho),
\end{align}
where $C_{k\ell m}$ are the expansion coefficients in Equation~(\ref{eq:bound_st_expansion}) and $J_{1,\kappa}^{\ell'm',\ell m}(\rho)$ is the angular integral with the volume element,
\begin{align}
J_{L,M}^{(\ell'm',\ell m)}(\rho) =& \int_{0}^{2\pi}\int_{-1}^{1}\rmd\eta\rmd\phi\, Y_{\ell'}^{m'*}(\eta,\phi) \sqrt{\tfrac{4\pi}{2L+1}}Y_{L}^{M}(\eta,\phi)Y_{\ell}^{m}(\eta,\phi) \left[ \rho(\rho+R)+\tfrac{R^{2}}{4}(1-\eta^{2}) \right].
\end{align}
The angular integral can be evaluated analytically
\begin{align}
J_{1,0}^{(\ell'm',\ell m)}(\rho) =&  \left[\rho(\rho+R)+\tfrac{R^{2}}{10} \right]A_{m'}^{\ell'}{}_{0}^{1}{}_{m}^{\ell} - \tfrac{R^{2}}{10} A_{m'}^{\ell'}{}_{0}^{3}{}_{m}^{\ell}, \\
J_{1,\pm1}^{(\ell'm',\ell m)}(\rho) =&  \left[\rho(\rho+R)+\tfrac{R^{2}}{5} \right]A_{m'}^{\ell'}{}_{\pm1}^{1}{}_{m}^{\ell} - \tfrac{R^{2}}{5\sqrt{6}} A_{m'}^{\ell'}{}_{\pm1}^{3}{}_{m}^{\ell},
\end{align}
where 
\begin{align}
A_{m_{1}}^{\ell_{1}}{}_{m_{2}}^{\ell_{2}}{}_{m_{3}}^{\ell_{3}} &= \int_{0}^{2\pi}\int_{-1}^{1} \!\rmd\eta\rmd\phi\, Y_{\ell_{1}}^{m_{1}*}(\eta,\phi) \sqrt{\tfrac{4\pi}{2\ell_{2}+1}} Y_{\ell_{2}}^{m_{2}}(\eta,\phi) Y_{\ell_{3}}^{m_{3}}(\eta,\phi) \nonumber \\
&= (-1)^{\ell_{2}}\,C_{\ell_{1}0\ell_{2}0}^{\ell_{3}0}\,C_{\ell_{3}m_{3}\ell_{2}m_{2}}^{\ell_{1}m_{1}}.
\end{align}
In the velocity gauge the components of the electronic dipole operator are
\begin{align}
\braket{\psi_{E\lambda}^{m'}|\mathsf{V}_{0}|\psi_{n}^{m}} &= \sum_{\ell'}\sum_{k\ell} \mathcal{C}_{\ell'm'}^{\lambda} C^{(n)}_{k\ell }\,A_{m'}^{\ell'}{}_{0}^{1}{}_{m}^{\ell} \int_{0}^{\infty}\!\rmd\rho\,\,  \Xi_{\lambda}^{\abs{m'}}(\rho) \!\left[ \rho(\rho+R)\frac{\partial}{\partial\rho} + y_{\ell'\ell}\left(\rho+\tfrac{R}{2} \right)\right]  f_{k}^{m}(\rho) , \\[7pt]
\braket{\psi_{E\lambda}^{m'}|\mathsf{V}_{\pm1}|\psi_{n}^{m}} &= -\sum_{\ell'}\sum_{k\ell} \mathcal{C}_{\ell'm'}^{\lambda}C^{(n)}_{k\ell }\,A_{m'}^{\ell'}{}_{\pm1}^{1}{}_{m}^{\ell}  \int_{0}^{\infty}\!\rmd\rho\,\,  \Xi_{\lambda}^{\abs{m'}}(\rho) 
 \sqrt{\rho(\rho+R)}\left[ \left(\rho+\tfrac{R}{2}\right)\!\frac{\partial}{\partial\rho} + \frac{\abs{m}R^{2}}{4\rho(\rho+R)} +y_{\ell'\ell}\right] f_{k}^{m}(\rho),
\end{align}
where
\begin{align}
y_{\ell'\ell} &= \begin{cases}
-\ell &\ell'=\ell+1, \\
\ell+1 &\ell'=\ell-1.
\end{cases}
\end{align}

\section{Origin of the Oscillations in the Vibrationally Resolved Cross Sections}
\label{App:osc}

\begin{figure}[!htb]
	\centering
	\includegraphics[width=0.6\linewidth]{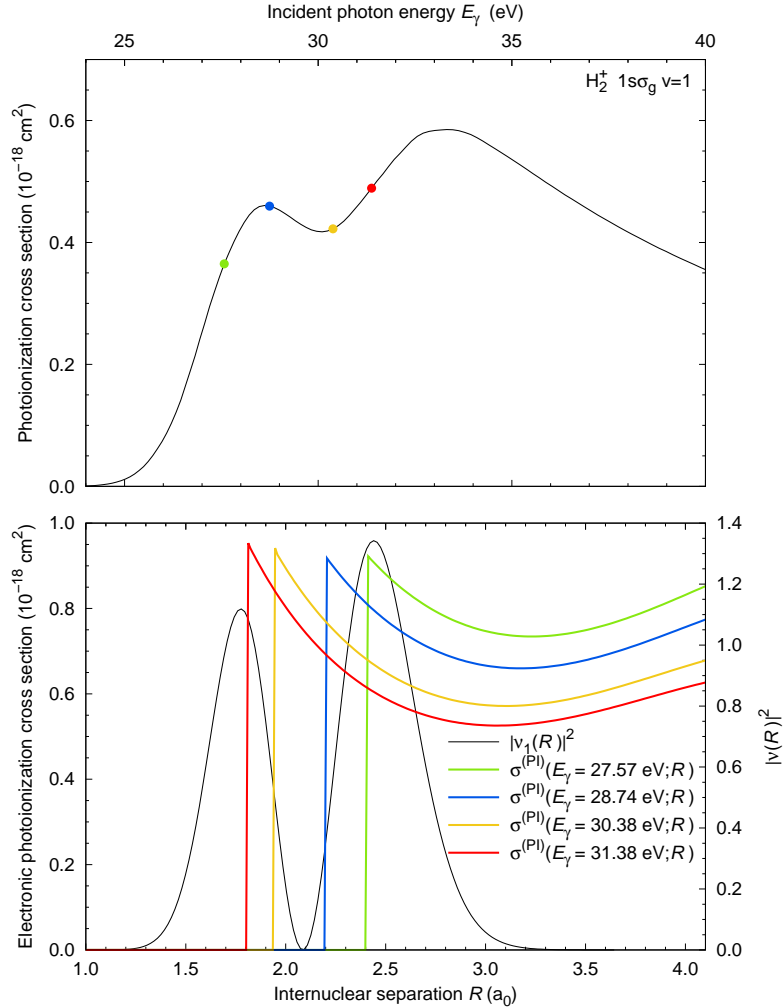}
	\caption{(Top) The near threshold $v=1$ photoionization cross section as a function of incident photon energy.
	(Bottom) Norm squared of the $v=1$ vibrational wave function and the electronic part of the photoionization cross section, Equation~(\ref{eq:FN_ion}), as a function of internuclear separation $R$. The first non--zero point of the electronic part of the photoionisation cross section shifts to a lower value of $R$ as the incident photon energy increases. The corresponding points in the top panel are indicated with dots of the same color.} \label{fig:vib_test}
\end{figure}

In Figure~\ref{fig:vib_test} the origin of these oscillations is investigated.
The top panel of Figure~\ref{fig:vib_test} shows the $v=1$ photoionization cross section in the region of the oscillations, while the bottom panel shows the components of the integrand in Equation~(\ref{eq:vib_ion}) used to calculate the cross section in the top panel.
It can be seen that the overlap of the squared $v=1$ vibrational wave function with the electronic part of the photoionization cross section, Equation~(\ref{eq:FN_ion}), increases as the incident photon energy increases until a node in the vibrational wave function is reached.
When a node is reached the overall overlap between the wave function squared and the cross section decreases. 
Therefore the near threshold oscillations in the photoionization cross sections for vibrationally excited states is due to nodes in the excited vibrational wave functions.

The shift of the ionization threshold to lower energies as $v$ increases can also be explained by the by the overlap between the electronic part of the photoionization cross section and the norm squared of the vibrational wave function.
The vibrational wave function extends to larger internuclear separations as the vibrational level increases.
It can been seen in Figure~\ref{fig:vib_test} at lower incident photon energies the first non--zero point of the electronic part of the photoionization cross section occurs at a larger internuclear separation.
Therefore, higher vibrational levels have wave functions that overlap with the electronic part of the photoionization cross section at lower incident photon energies, this results in the shift of the ionization threshold to lower energies.

\pagebreak

\bibliography{references}
\bibliographystyle{aasjournal}



\end{document}